# Exploring the next step in micro-solvation of CO in water: Infrared spectra and structural calculations of $(H_2O)_4$- CO and $(D_2O)_4$- CO

A.J. Barclay,[a] A. Pietropolli Charmet,[b] A.R.W. McKellar,[c] and N. Moazzen-Ahmadi[a]

[a] *Department of Physics and Astronomy, University of Calgary, 2500 University Drive North West, Calgary, Alberta T2N 1N4, Canada*

[b] *Dipartimento di Scienze Molecolari e Nanosistemi, Università Ca' Foscari Venezia, Via Torino 155, I-30172, Mestre, Venezia, Italy*

[c] *National Research Council of Canada, Ottawa, Ontario K1A 0R6, Canada*

**Abstract**

We extend studies of micro-solvation of carbon monoxide by a combination of high-resolution IR spectroscopy and *ab initio* calculations. Spectra of the $(H_2O)_4$-CO and $(D_2O)_4$-CO pentamers are observed in the C-O stretch fundamental region ($\approx 2150$ cm$^{-1}$). The $H_2O$ containing spectrum is broadened by predissociation, but that of $D_2O$ is sharp, enabling detailed analysis which gives a precise band origin and rotational parameters. *Ab initio* calculations are employed to confirm the assignment to (water)$_4$-CO and to determine the structure, in which the geometry of the (water)$_4$ fragment is a cyclic ring very similar to the isolated water tetramer. The CO fragment is located "above" the ring plane, with a partial hydrogen bond between the C atom and one of the "free" protons (deuterons) of the water tetramer. Together with previous results on $D_2O$-CO, $(D_2O)_2$-CO, and $(D_2O)_3$-CO, this represents a probe of the four initial steps in the solvation of carbon monoxide at high resolution.

**1. Introduction**

The study of high resolution spectra of weakly-bound molecular clusters becomes increasingly challenging as cluster size increases, just as high-level quantum chemical calculations of cluster structures and potential surfaces also become more difficult. In a recent paper,[1] we analyzed infrared spectra of the $(D_2O)_2$-CO and $(D_2O)_3$-CO clusters, together with *ab initio* calculations which helped to confirm their most stable structures. Along with the known dimer, $D_2O$-CO, these clusters represented the first three steps in the solvation of carbon monoxide in water. In the present paper, we extend these spectroscopic and quantum chemical studies to the $(D_2O)_4$-CO pentamer. Theory and experiment are complementary, since the quantum calculations offer important help in the interpretation of the experimental results, and the analyzed spectra can then help to verify and refine the calculations.

The very first solvation step for carbon monoxide is of course the water-CO dimer, known to have a planar equilibrium structure with almost co-linear heavy atoms (O, C, O) and a hydrogen bond between the water and the carbon atom. Interchange of H (or D) atoms gives rise to two spectroscopically resolved tunneling states, which correspond to different nuclear spin species. For the isotopomer with HDO, there is no tunneling but instead there is a deuteron bound isomer HOD-CO, and a proton bound isomer DOH-CO. Detailed studies of $H_2O$-CO, $D_2O$-CO, HOD-CO, and DOH-CO are available in the microwave,[2] millimeter wave,[3] and infrared[4-8] regions. High-level *ab initio* calculations are available for the binary water-CO potential energy surface,[7,9-11] together with many for water-water.[12] However, in this and our previous paper[1] we make new calculations directly for the larger water-CO clusters themselves, rather than assuming additivity of two body forces.

## 2. Experimental spectra and analysis

Spectra were recorded as described previously[1,8,13-15] with a pulsed supersonic slit jet expansion probed by a rapid-scan tunable infrared quantum cascade laser. The gas mixture contained about 0.02 to 0.06% CO in helium carrier gas, with a backing pressure of about 10 atmospheres. $H_2O$ or $D_2O$ was added using a 'bubbler' in the gas line to the jet, and we estimate the resulting water concentration to be about 0.01%, or perhaps a bit more. Although (water)$_n$-CO, n=1,2,3 bands could be observed using water and CO in a mixing cylinder, the bubbler was essential for observing the present spectrum. Wavenumber calibration was achieved by recording signals from a fixed etalon and a reference gas cell containing room temperature $N_2O$ gas. Spectral simulations were made using the powerful PGOPHER software package.[16]

The infrared bands of $(D_2O)_2$-CO and $(D_2O)_3$-CO, which we reported previously,[1] are located at 2158.40 and 2149.65 cm$^{-1}$, respectively. For comparison, the $D_2O$-CO band center is 2154.54 cm$^{-1}$, and that of the free CO monomer is 2143.271 cm$^{-1}$. The corresponding $(H_2O)_2$-CO and $(H_2O)_3$-CO cluster spectra were also observed, but they were broadened by predissociation in the upper state (shortened lifetimes) which made precise rotational analysis impossible. The observed magnitude of this broadening was roughly 0.009 cm$^{-1}$ (FWHM), corresponding to a predissociation lifetime around 0.6 ns. The band origins for the $H_2O$ clusters were 0.8 to 0.9 cm$^{-1}$ lower than those of the corresponding $D_2O$ clusters. For the water-CO dimer, in contrast, reasonably sharp (<0.0015 cm$^{-1}$) spectra are observed for all species ($H_2O$-, HOD-, and $D_2O$-CO) in the CO fundamental region, giving lifetimes of at least 3 ns.

The new bands reported here are centered around 2144.1 cm$^{-1}$ and were observed using experimental conditions similar to those for the smaller clusters with the exception that a bubbler in the gas line was necessary.[1] They are shown in Figs. 1 and 2, where short gaps in the observed spectra correspond to known lines[15] of CO and $(CO)_2$, which are clipped out for clarity. As was the case for the smaller clusters,[1] we observe a sharp spectrum using $D_2O$ and a broadened one using $H_2O$. Rotational analysis of the $D_2O$ spectrum was more challenging than for $(D_2O)_3$-CO, and $(D_2O)_2$-CO, even though in some ways the band structure appeared simpler, particularly in the central region (Fig. 2). Simplicity suggested a smaller cluster, but various possibilities did not work out, including, for example, the known[17] water-$(CO)_2$ trimer. Eventually, we realized that the pattern of the $D_2O$-containing spectrum was close to that of a symmetric top (helping to explain the simplicity) and that this near symmetric top had rotational constants which were consistent with a larger cluster, namely $(D_2O)_4$-CO.

The parameters listed in Table I provide a remarkably accurate representation of the $D_2O$ containing spectrum, as shown by the simulations in the lower panel of Fig. 1, and especially in the expanded view in Fig. 2. We assigned 105 observed lines from the $D_2O$ spectrum in terms of 274 transitions, and fitted them with an rms error of about 0.00041 cm$^{-1}$. (There are many blended lines, including unresolved asymmetry doublets). The excellent interactive tools of PGOPHER[16] aided the assignment and simulation, and we used the mergeblend option, fitting blended lines to an intensity weighted average of their calculated components. In addition to the rotational parameters in Table I, further structural information can be inferred from the relative strength of *a*-, *b*-, and *c*-type rotational transitions, since this depends on the orientation of the CO monomer within the cluster. We estimated these relative intensities to be roughly 0.14 : 1.0 : <0.01, respectively. In other words, the band is mostly *b*-type, with a smaller but significant *a*-type component (this accounts for the central *Q*-branch peak at 2144.124 cm$^{-1}$), and a negligible *c*-type component.

The spectrum obtained using $H_2O$ (upper panel of Fig. 1) was significantly broadened, just as found for $(H_2O)_2$-CO and $(H_2O)_3$-CO,[1] so that detailed rotational analysis was not possible. The magnitude of the broadening was roughly 0.012 cm$^{-1}$ (FWHM), corresponding to a predissociation lifetime of about 0.5 ns. The $H_2O$ species band origin is 2144.05 cm$^{-1}$, which is only 0.07 cm$^{-1}$ lower than that of the $D_2O$ species.

We suspected that the species responsible for the new cluster band was $(D_2O)_4$-CO based on the fitted rotational parameters (Table I) and on the fact that $(D_2O)_2$-CO and $(D_2O)_3$-CO were previously observed[1] under similar conditions. But there are many possible ways to arrange four water and one carbon monoxide molecules. Moreover, our tentative assignment to $(D_2O)_4$-CO could be wrong. So we turned to theoretical structure calculations for further guidance.

### 3. Computational details

Following our previous studies on molecular clusters,[18,19] the potential energy surface (PES) of $(D_2O)_4$-CO was characterized at DFT level of theory in conjunction with the m-aug-cc-pVTZ-dH basis set[20] (-dH means that polarization functions have been removed for the hydrogen atom). Dispersion effects[21,22] were taken into account for all the DFT functionals here employed by using the D3BJ corrections.[23,24] Since there are three most stable isomers for $(D_2O)_4$ (they are labeled as CY4-1, CY4-2 and P4),[25] for each one of them 150 different starting geometries for $(D_2O)_4$-CO were generated by randomly adding the CO molecule; to achieve a better sampling of the PES, an additional set of 150 geometries were originated by randomly distributing in space four $D_2O$ molecules and one of CO. Thus, a total of 600 initial structures were created. Then, each one was fully optimized (at B3LYP level of theory),[26,27] and subsequent hessian calculations were performed to select only the ones corresponding to true minima. These structures were then grouped according to their equilibrium rotational constants. The geometries of the true minima were further refined by using the B2PLYP functional.[28] We thus found nine possible low-lying isomers for $(D_2O)_4$-CO. Appropriate extrapolation methods to the complete basis set (CBS) limit make it possible to account for both basis set incompleteness (BSIE) and superposition (BSSE) errors when predicting different molecular properties (see, for example, Refs. 29, 30, and references therein). For the accurate determination of the energetics of all the structures investigated in the present work, we used the composite schemes based on the three-point extrapolation at CBS limit proposed by Feller.[31] The corresponding results for the binding energies are reported in Table II. Vibrational corrections to these equilibrium rotational constants were calculated by second order vibrational perturbation theory (VPT2) and employing the B2PLYP functional, given its good results for the cubic part of the potential.[32] The potential bias due to intermolecular motions was accounted for by using a reduced-dimensionality scheme and an appropriate suite of programs.[33]

The last column of Table I reports the computed rotational parameters for geometry #2, the most stable calculated isomer of $(D_2O)_4$-CO. The agreement with experiment is quite satisfactory, and confirms the assignment to $(D_2O)_4$-CO. The corresponding basic equilibrium cluster geometry (without vibrational corrections) is given in Table III and illustrated in Fig. 3. Structures for the other possible isomers in Table II are given in the supplementary material given below. They not only had higher calculated energies but also had incompatible rotational constants. Besides, by taking the CBS binding energy for the most stable structure and the ones we previously obtained for $(D_2O)_n$-CO ($n$ = 2 and 3) clusters,[1] it is possible to derive for each of them the energy required for addition of CO to the corresponding $D_2O$ dimer, trimer and tetramer, according to the reaction, $(D_2O)_n$ + CO → $(D_2O)_n$-CO, for $n$ = 2, 3 and 4. This reaction is exothermic, with values of -4.1, -6.0 and -7.5 kcal/mol per $D_2O$ molecule for the structures containing the water dimer, trimer and tetramer, respectively; this trend is in line with the one previously determined for the addition of $CO_2$ to water clusters.[34]

In a similar way, it is possible to derive also the values for the addition of one $D_2O$ molecule to a $(D_2O)_n$-CO cluster to yield the corresponding $(D_2O)_{n+1}$-CO structure, according to the reaction $(D_2O)_n + CO \rightarrow (D_2O)_{n+1}$-CO, for $n = 2$ and 3. This reaction is also exothermic, with values of -9.8 and -11.8 kcal/mol for the addition of one water molecule to the clusters containing the water dimer and the trimer, respectively; again these values are similar to those obtained when considering the similar clusters between water and $CO_2$.[34]

## 4. Discussion and conclusions

The simulated spectrum for $(D_2O)_4$-CO in Figs. 1 and 2 is based on the experimental parameters from Table I, with an effective rotational temperature of 2 K, and an effective instrumental resolution of 0.0025 cm$^{-1}$. To estimate a simulated spectrum for $(H_2O)_4$-CO, we scaled the experimental $(D_2O)_4$-CO rotational parameters by the ratios of the calculated parameters for $(H_2O)_4$-CO and $(D_2O)_4$-CO (without vibrational corrections). After convolution with a Lorentzian of width 0.012 cm$^{-1}$, the result provides a good qualitative match to experiment (top panel of Fig. 1). There is, however, a region of unexplained extra experimental absorption around 2143.9 to 2144.1 cm$^{-1}$. We think this is unrelated to $(H_2O)_4$-CO and due to a different, possibly larger, cluster. Another broad unresolved feature, not shown here, appears in the $H_2O$ spectrum around 2145.4 to 2145.5 cm$^{-1}$.

The structure of the water tetramer fragment within our cluster (Fig. 3) is only slightly distorted from that of the isolated water tetramer itself, which has a highly symmetric ($S_4$ point group) hydrogen-bonded cyclic ring structure.[35] It seems that the addition of CO does not greatly affect the hydrogen bonds ruling the structure of the water tetramer, consistent with the results previously obtained for both $(D_2O)_2$-CO and $(D_2O)_3$-CO. In the present case, we observe that each water molecule has one proton (or deuteron) involved in a nearly linear hydrogen bond with its neighbor, while the other proton sticks out from the plane formed by the four O atoms. These out-of-plane protons alternate on one side or the other. Isolated $(D_2O)_4$, which is a symmetric rotor, has a measured $A$ rotational constant of 3080 MHz,[36] while the slightly distorted $(D_2O)_4$ fragment in our calculated equilibrium cluster is an oblate near symmetric rotor with $A = 3187$ and $B = 3137$ MHz. In the isolated tetramer the distance between neighboring O atoms is 2.74 Å, while in (water)$_4$-CO, this length varies from 2.73 to 2.75 Å.

The CO fragment in the calculated equilibrium cluster has its center of mass located 3.08 Å "above" the near-plane defined by the four water O atoms, but displaced "sideways" by 0.33 Å from the near-symmetry axis which is perpendicular to the plane. (This is also close to the *a*-inertial axis of the cluster.) The CO orientation suggests a partial hydrogen bond between the C atom and one of the deuterons that points out of the plane in this direction (this C-D distance is 2.85 Å) and perhaps another weak H-bond between the O atom (of the CO) and the other deuteron pointing out of the plane in this direction (O-D distance 3.13 Å). The CO molecular axis makes angles of 79.3°, 11.4°, and 86.2° with the cluster *a*-, *b*-, and *c*-axes, respectively, predicting a dominant *b*-type spectrum and a negligible *c*-type contribution, just as observed. But the predicted *a*-type relative intensity (0.036) is significantly smaller than observed (≈0.14), suggesting that the CO molecule in the cluster should be rotated in the *ab*-plane so that the effective *a*- and *b*-axis angles are more like 70° and 20°, rather than 79° and 11°. In other words, the true structure evidently has the CO pointing more towards the plane of the $(D_2O)_4$ fragment than shown in Fig. 3.

The measured $(D_2O)_4$-CO band origin represents a vibrational shift of +0.85 cm$^{-1}$ relative to the free CO molecule. For comparison, much larger shifts were observed[1,4] for $(D_2O)_3$-CO (+6.38 cm$^{-1}$), $(D_2O)_2$-CO (+15.12 cm$^{-1}$), and $D_2O$-CO (+11.27 cm$^{-1}$). We note that these shifts are inversely correlated (very roughly!) with the respective C – D hydrogen bond lengths, which are: $(D_2O)_4$-CO (2.85 Å), $(D_2O)_3$-CO (2.73 Å),[1] $(D_2O)_2$-CO (2.30 Å),[1] and $D_2O$-CO (2.36 Å).[10] These quoted bond lengths are calculated equilibrium values, so they neglect large amplitude motion effects which may be especially significant for $(D_2O)_2$-CO and $D_2O$-CO. Nevertheless, we suspect that the relatively small vibrational shift for $(D_2O)_4$-CO is related to a weaker C – D hydrogen bond.

In conclusion, the weakly-bound $(H_2O)_4$-CO and $(D_2O)_4$-CO clusters have been studied experimentally by high-resolution infrared spectroscopy in the C-O stretch fundamental region, and theoretically by means of *ab initio* calculations. As previously observed for the (water)$_3$-CO and (water)$_2$-CO clusters, the $H_2O$ containing spectrum was broadened (≈0.012 cm$^{-1}$) by predissociation, while the $D_2O$ spectrum was sharp, allowing detailed rotational analysis The well determined rotational parameters (*A*, *B*, *C*) from the analysis are not sufficient to determine a cluster structure, but the *ab initio* calculations give parameters for the lowest energy $(D_2O)_4$-CO isomer in good agreement with experiment, providing confirmation of the assignment to $(D_2O)_4$-CO. In the resulting cluster structure, the $(D_2O)_4$ fragment is a cyclic hydrogen-bonded ring, very similar to the isolated $(D_2O)_4$ tetramer. This is not so surprising, since water tetramer is noted to be particularly stable, thanks in part to its high symmetry.[35,36] The CO molecule lies "above" the plane of the water tetramer ring, with a C – D hydrogen bond involving one of the "free" deuterons of the $(D_2O)_4$ fragment. Relative intensities of observed transitions in the spectrum suggest that in the true effective structure, the CO is somewhat rotated so that the C atom points more toward the ring plane.

**Acknowledgements**

The financial support of the Natural Sciences and Engineering Research Council of Canada is gratefully acknowledged. We thank K. Michaelian for the loan of the QCL.

**Data availability**

The data that support the findings of this study are available from the corresponding author upon reasonable request.

Table I. Experimental and theoretical spectroscopic parameters for $(D_2O)_4$-CO. [a]

|  | Experiment | | Theory |
|---|---|---|---|
|  | Excited state | Ground state | Ground state |
| $\nu_0$ / cm$^{-1}$ | 2144.1234(1) |  |  |
| $\Delta\nu_0$ / cm$^{-1}$ | +0.852 |  |  |
| $A$ / MHz | 1569.50(33) | 1569.96(30) | 1582.1 |
| ½$(B + C)$ / MHz | 1347.54(33) | 1347.07(28) | 1359.1 |
| $(B - C)$ / MHz | 43.29(57) | 45.21(51) | 64.2 |
| $D_J$ / kHz | 1.3(21) | 3.8(15) |  |

[a] Numbers in parentheses are 1σ uncertainties in units of the last quoted digit. $\Delta\nu_0$ is the vibrational shift relative to the free CO molecule. Theoretical values are equilibrium rotational constants augmented by vibrational corrections (see text).

Table II. CBS extrapolated binding energies (BE, in kcal/mol) for the several $(D_2O)_4$-CO isomers identified. [a]

| Geometry | BE |
|---|---|
| 1 | -29.81 (-22.65) |
| 2 | -29.80 (-22.89) |
| 3 | -29.80 (-22.79) |
| 4 | -28.73 (-21.87) |
| 5 | -28.73 (-21.87) |
| 6 | -28.69 (-21.69) |
| 7 | -28.66 (-21.68) |
| 8 | -28.55 (-21.75) |
| 9 | -28.28 (-21.59) |

[a] Values in parentheses include zero-point vibrational energy.

Table III. Calculated structure for $(D_2O)_4$-CO.[a]

| Atom | a | b | c |
| --- | --- | --- | --- |
| C | 2.0885390 | 0.7452570 | 0.5751264 |
| O | 2.2995834 | -0.3636982 | 0.6498239 |
| O | -0.9086208 | 1.9084153 | -0.1516765 |
| D | -0.6247902 | 1.3595166 | -0.9124168 |
| D | -0.1406856 | 2.4413274 | 0.0721530 |
| D | -1.3388174 | 0.6795682 | 1.0396037 |
| O | -1.4779548 | -0.1380985 | 1.5657432 |
| D | -2.3713226 | -0.0837602 | 1.9131247 |
| O | -0.0660099 | 0.0604856 | -2.0020124 |
| D | -0.4593767 | -0.0766911 | -2.8669857 |
| D | -0.2385661 | -0.7592986 | -1.4900600 |
| O | -0.6381580 | -1.9790127 | -0.2778089 |
| D | 0.0811140 | -2.4950081 | 0.0947890 |
| D | -0.9868406 | -1.4356410 | 0.4605737 |

[a]These are principal axis system Cartesian coordinates in units of Angstroms.

## Figure Captions

Fig. 1. Observed and simulated spectra of the $(H_2O)_4$-CO and $(D_2O)_4$-CO clusters in the region of the CO stretch vibration. The $H_2O$ spectrum is broadened by predissociation, but the $D_2O$ spectrum is sharp, allowing detailed rotational analysis. The simulated trace for $(D_2O)_4$-CO uses the parameters of Table I and an effective rotational temperature of 2 K. The simulated trace for $(H_2O)_4$-CO uses estimated parameters, scaled from those of $(D_2O)_4$-CO (see text). For clarity of presentation, there are small gaps in the observed traces where known lines of CO monomer and CO dimer have been "clipped out".

Fig. 2. Central section of the $(D_2O)_4$-CO spectrum. The simulated spectrum here and in Fig. 1 consists almost entirely of *b*-type transitions, except for the unresolved *a*-type *Q*-branch at 2144.12 cm$^{-1}$.

Fig. 3 Calculated equilibrium structure of $(H_2O)_4$-CO as given in Table III. The upper panel shows a "side" view along the *c*-inertial axis of the cluster, with the CO fragment located "above" the near-plane defined by the four water O atoms. The lower panel shows a view from "above" along the cluster *a*-axis, with the CO fragment in the foreground. In the true structure, the CO probably should be rotated a bit in the *ab* plane so that it has a larger projection on the cluster *a*-axis (see text).

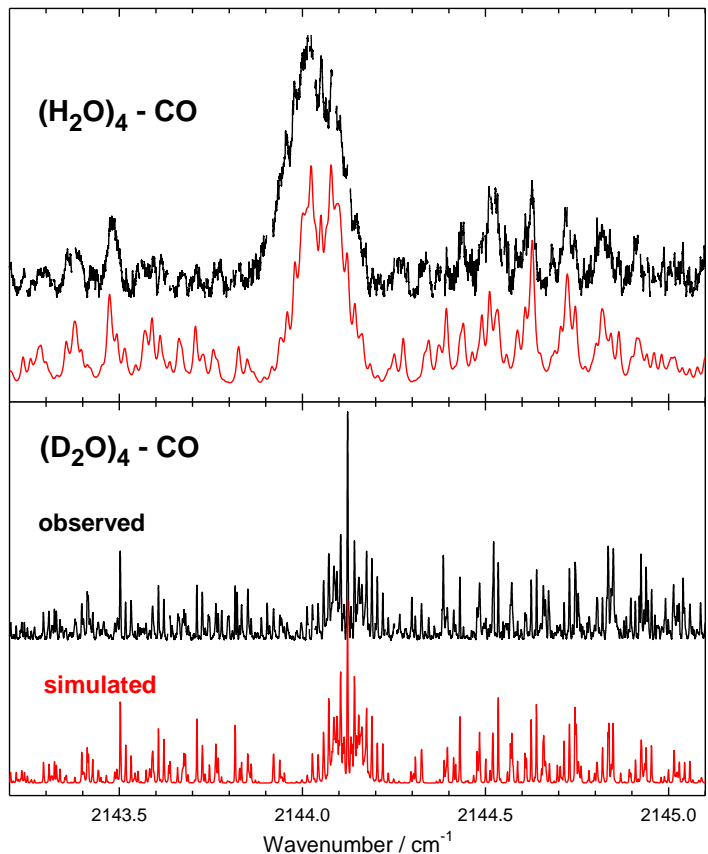

Fig. 1. Observed and simulated spectra of the $(H_2O)_4$-CO and $(D_2O)_4$-CO clusters in the region of the CO stretch vibration. The $H_2O$ spectrum is broadened by predissociation, but the $D_2O$ spectrum is sharp, allowing detailed rotational analysis. The simulated trace for $(D_2O)_4$-CO uses the parameters of Table I and an effective rotational temperature of 2 K. The simulated trace for $(H_2O)_4$-CO uses estimated parameters, scaled from those of $(D_2O)_4$-CO (see text). For clarity of presentation, there are small gaps in the observed traces where known lines of CO monomer and CO dimer have been "clipped out".

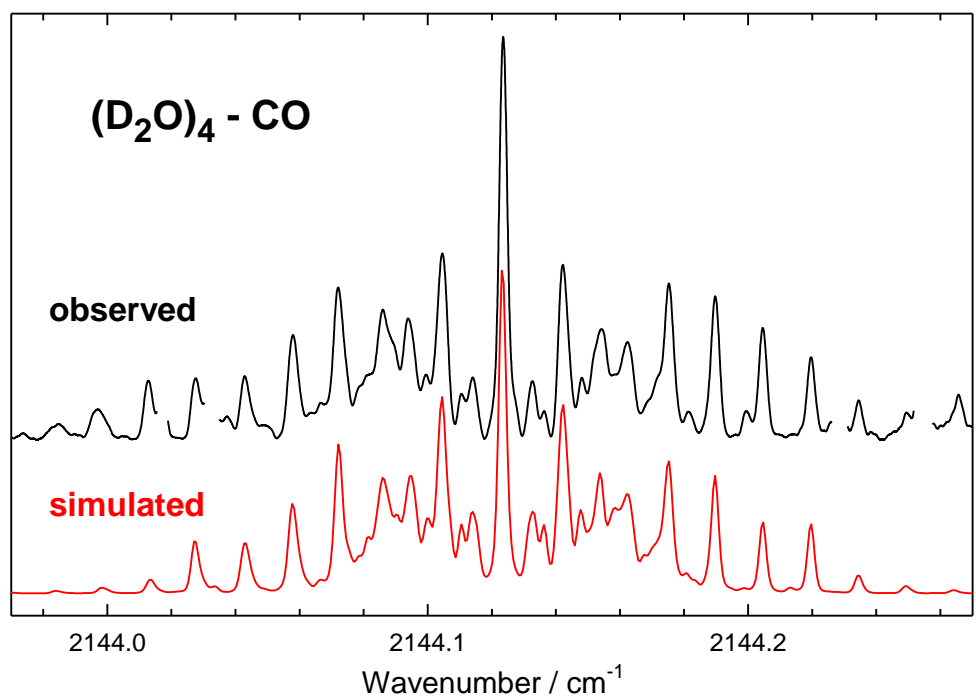

Fig. 2. Central section of the $(D_2O)_4$-CO spectrum. The simulated spectrum here and in Fig. 1 consists almost entirely of *b*-type transitions, except for the unresolved *a*-type *Q*-branch at 2144.12 cm$^{-1}$.

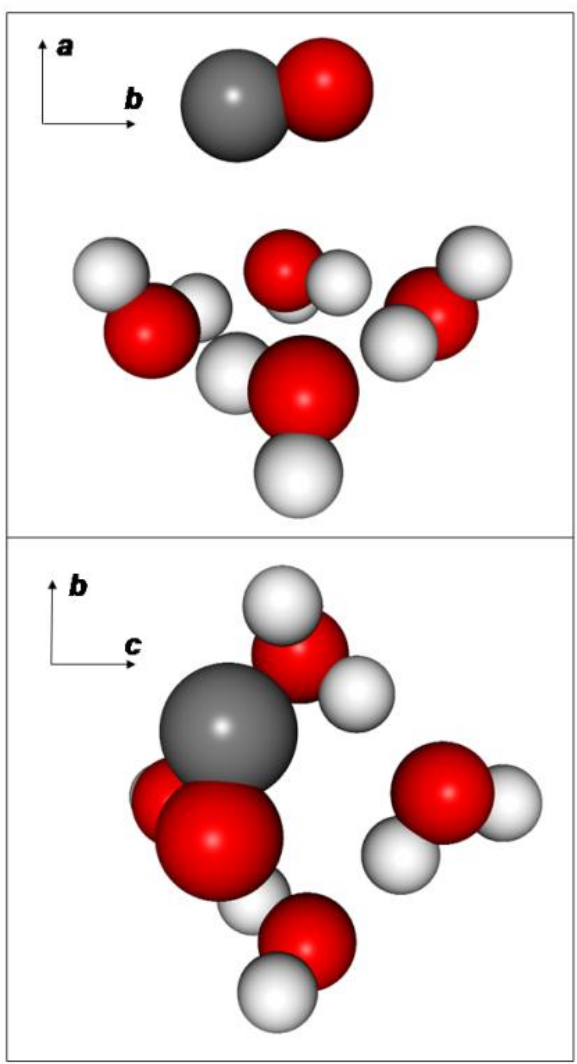

Fig. 3. Calculated equilibrium structure of $(H_2O)_4$-CO as given in Table III. The upper panel shows a "side" view along the $c$-inertial axis of the cluster, with the CO fragment located "above" the near-plane defined by the four water O atoms. The lower panel shows a view from "above" along the cluster $a$-axis, with the CO fragment in the foreground. In the true structure, the CO probably should be rotated a bit in the $ab$ plane so that it has a larger projection on the cluster $a$-axis (see text).

## SUPPLEMENTARY MATERIAL

**COMPUTATIONAL METHODOLOGY**

For computing the CBS limit of the binding energies, many single point energy calculations were performed on the optimized geometries (obtained at B2PLYP-D3BJ level of theory) in conjunction with the aug-cc-pV$n$Z basis sets with $n$ = T, Q and 5 were employed.[1,2] Besides, some calculations were done using the cc-pCVTZ basis set.[3]

We followed this composite scheme, which is based on separate extrapolation of the CBS limit for the Hartree-Fock (HF-SCF) energy, $E_{CBS}$(HF-SCF) and for the correlation energy computed at MP2 level of theory, $E_{CBS}(corr)$. For computing the $E_{CBS}$(HF-SCF) we used the expression proposed by Fenner[4], while for $E_{CBS}(corr)$ we employed the following inverse cubic function

$$E_{CBS}(corr) = \frac{N^3 E_N^{corr} - (N-1)^3 E_{N-1}^{corr}}{N^3 - (N-1)^3}.$$

The CBS limit was then obtained as:

$$E_{CBS} = E_{CBS}(\text{HF} - \text{SCF}) + E_{CBS}(corr)$$

The effects due to higher-order electron correlation past the MP2 level of theory were accounted for as the difference between the CCSD(T) and MP2 single point energies, calculated using the aug-cc-pVTZ basis set.

$$\Delta_{MP2}^{CCSD(T)} = E_{CCSD(T)} - E_{MP2}$$

For computing the core-valence (CV) corrections, we performed two calculations at MP2 level and using the cc-pCVTZ basis set, correlating all the electrons, $E_{ae-MP2}$, and within the frozen-core approximation, $E_{fc-MP2}$.

$$\delta_{MP2}^{CV} = E_{ae-MP2} - E_{fc-MP2}$$

The CBS-1 limit was therefore given as

$$E_{CBS-1} = E_{CBS}(\text{HF} - \text{SCF}) + E_{CBS}(corr) + \Delta_{MP2}^{CCSD(T)} + \delta_{MP2}^{CV}$$

By using the data thus obtained (and with the inclusion of zero-point vibrational correction), the most stable isomer of $(D_2O)_4$-CO among the different structures optimized was therefore identified. The DFT-D3BJ calculations were carried out using the ORCA suite of programs,[5] single point energy calculations were carried out using the Gaussian suite of quantum chemical programs.[6] For the geometry optimizations the *TightOpt* criteria, as implemented in the Orca software, were employed.

******************************************************************************
Optimized geometries (Angstrom) for (D$_2$O)$_4$-CO at B2PLYP-D3BJ/m-aug-cc-pVTZ-dH level of theory

Geo #1
O  -1.63818497439361    0.78450170942506    0.18960086176552
D  -1.50077952480934   -0.18487578517861    0.13122586703526
D  -2.31648849249492    0.99277465605407   -0.45717830896441
D   0.03322838716825    1.40736064272081    0.00017222379479
O   1.00038359052084    1.52964301957340   -0.09110263875635
D   1.27839619778856    2.03408087273902    0.68006882573315
O  -0.88640812198796   -1.84020612319305   -0.06049363825853
D  -0.95161454321030   -2.47425762242800    0.65746484270175
D   0.07365595997946   -1.69865532286994   -0.20859623116477
O   1.71649487002351   -1.07604717598203   -0.40014507086530
D   2.18517164319315   -1.19187145009491   -1.23001773032812
D   1.58432898749369   -0.10693959496095   -0.29743052360147
C   2.03096243282128    3.15235233252727    2.57502456749823
O   2.39085357790738    3.67213982166783    3.51140695341023

Geo #2
```
C   2.07920086798628   -0.94066517791645   -0.26824242708418
O   2.39877377685638    0.14401698203279   -0.30344035962624
O  -1.08396258848902   -1.82181426502063    0.01832253867644
D  -0.86181772425673   -1.29736360836817    0.81593053764601
D  -0.34382990676734   -2.42369033907017   -0.09901564707327
D  -1.22761942785676   -0.56472718299936   -1.21187160827771
O  -1.21589624882164    0.25961878691517   -1.74579780050961
D  -2.05233690433454    0.28533147042113   -2.21637033382419
O  -0.34686391635747   -0.04939243200276    1.98397419436813
D  -0.84449681041238    0.12738238280454    2.78575093817275
D  -0.36989747108693    0.78018232607894    1.45938655267184
O  -0.48125068485028    2.02527784514654    0.21268429407082
D   0.32745987423737    2.47162572577517   -0.05036223487114
D  -0.76998283584692    1.51236048620327   -0.57191164433965
```

Geo #3
```
O  -1.09427793076395   -1.37463264140794   -0.07740767954757
D  -0.11501163024293   -1.34824590145076   -0.00036735790340
D  -1.28873977016773   -2.00276115535048   -0.77694105282506
D  -1.51846878670777    0.33237839659393   -0.22381883500269
O  -1.53961241957901    1.31291340096083   -0.26660025013403
D  -2.16059872807359    1.59401539559541    0.40945903831902
O   1.59136457104923   -0.92319355472328    0.09597511055045
D   2.01781999854355   -1.01206762404931    0.95340428506090
D   1.56404546931568    0.04282998283659   -0.06052903610185
O   1.14970830983481    1.78970783860352   -0.15027637031070
D   1.42704389470911    2.33785599798906   -0.88824548112409
D   0.16957655764241    1.75844167537366   -0.19075560647815
C   2.41920579790614    0.75684318168858    2.80631043584739
O   2.37794467653405    1.73591499734018    3.36979279964977
```
****************************************************************************
Optimized geometries (Angstrom) for $(D_2O)_4$-CO at B2PLYP-D3BJ/m-aug-cc-pVTZ-dH level of theory

Geo #4
```
C  -1.13202182737746   -0.81238101033058    0.58372293601609
O  -0.30795001739748   -0.03989818406769    0.54292789555701
D  -2.20155442612692   -5.10684085840026   -2.29346199018765
O  -2.57243742450019   -4.33885821123687   -2.73373951841390
D  -1.59495757347115   -3.19719582671662   -3.66297970603276
D  -2.96487518991172   -3.78060559458179   -2.02752939834708
O  -1.15796920780192   -2.41437735935658   -4.06178245823958
D  -1.22623156709923   -2.53005242299663   -5.01223369917770
D  -1.87419650978993   -1.04690147161861   -3.16700342364622
O  -2.33189194830125   -0.47633064949008   -2.51392208570454
D  -2.78700525007182    0.19932091092955   -3.02182823105790
```

| | | |
|---|---|---|
| D | -3.22239631820873 | -1.67296278483453 | -1.50637921825311 |
| O | -3.52397487577064 | -2.44869633962593 | -0.99138174666209 |
| D | -3.10253789417152 | -2.33422019767337 | -0.13440934585054 |

Geo #5

| | | | |
|---|---|---|---|
| C | 0.17741942954958 | 0.79483757542502 | -0.03735390920533 |
| O | 0.38790014125170 | -0.25831082798557 | -0.38956408035058 |
| D | 2.85268731656160 | 2.00246079190216 | 3.94627234894837 |
| O | 3.46948989192077 | 1.29226067335135 | 3.66716412531973 |
| D | 0.53145887373699 | 3.16287658127681 | 0.97133013476949 |
| D | 4.25469743897710 | 1.39891414186165 | 4.20896929233962 |
| O | 1.31393756498325 | 3.55482092948310 | 1.37022576299280 |
| D | 2.03897699784079 | 2.95147630176593 | 1.10861161523166 |
| D | 1.46829448766725 | 3.52673929253749 | 3.13990064564441 |
| O | 1.74075894080357 | 3.37053454504963 | 4.07034913536848 |
| D | 0.93229693799334 | 3.34892775517640 | 4.58715084106265 |
| D | 4.10221074843859 | 1.78173650921384 | 0.51508899972406 |
| O | 3.25987704963198 | 1.64077405069814 | 0.95350081363744 |
| D | 3.46999419064346 | 1.43195167024401 | 1.88835428451717 |

Geo #6

| | | | |
|---|---|---|---|
| C | 0.62996959582129 | 0.06156575361798 | -0.90955999581457 |
| O | -0.06445583539181 | -0.24839128977275 | -1.7451299498071 |
| D | 2.88688661859789 | 2.96734088097236 | 1.30296519157410 |
| O | 2.90280889199064 | 3.93456101784255 | 1.47910312900193 |
| D | 1.23099271946221 | 0.73847499488388 | 4.34942116803518 |
| D | 3.74892472018150 | 4.24710503146191 | 1.15073629349258 |
| O | 2.04638294573162 | 1.08104491296906 | 3.97652826849191 |
| D | 2.03601889131642 | 2.04611407564508 | 4.14981084127552 |
| D | 2.48218063725283 | 1.01189384900423 | 2.22001456756337 |
| O | 2.76345271597434 | 1.22103324137449 | 1.30670164530761 |
| D | 2.11386927847568 | 0.81370316358287 | 0.72515007864133 |
| D | 2.69648216770954 | 4.29444651064900 | 4.69433437605384 |
| O | 2.10558877288550 | 3.82069923607158 | 4.10452616665351 |
| D | 2.42089787999231 | 4.01040861169774 | 3.19539820953085 |

*********************************************************************************

Optimized geometries (Angstrom) for $(D_2O)_4$-CO at B2PLYP-D3BJ/m-aug-cc-pVTZ-dH level of theory

Geo #7

| | | | |
|---|---|---|---|
| C | -0.81282455532802 | 1.08694824817435 | -0.10833960170486 |
| O | -0.63470663276061 | 2.15075454737177 | 0.22774499851589 |
| D | -3.66980960813687 | -3.54214757458237 | -2.88366799548683 |
| O | -3.56052647542068 | -3.49935316511830 | -3.85812468343655 |
| D | -1.10630435692184 | -1.10948832444421 | -0.84878875457685 |
| D | -3.72038453042464 | -4.38948122958478 | -4.17983455573924 |
| O | -1.21677848495386 | -2.01635972320159 | -1.15091583272966 |
| D | -1.08233083985023 | -1.99428887992688 | -2.12032724905445 |
| D | -2.73306794483583 | -2.91018873181538 | -0.98865038277053 |
| O | -3.54567977800030 | -3.44530682717978 | -1.12427414088320 |
| D | -4.23616110242163 | -3.01315367859915 | -0.61650494727120 |
| D | -0.49667389570461 | -2.57494659548701 | -4.39727566926432 |
| O | -1.18147035496535 | -2.11328012747393 | -3.90790425498815 |
| D | -2.00328144027549 | -2.62970792813270 | -4.04313691061003 |

Geo #8

| | | | |
|---|---|---|---|
| C | 0.61600205723008 | 0.34648486542185 | 0.82612381266079 |
| O | 0.68504498969410 | 0.39949845080459 | -0.30066948924175 |
| D | 3.08023915299415 | 4.21687048250424 | 0.56366288720586 |
| O | 2.52145339992812 | 3.56861508494717 | 0.99810237529945 |
| D | 1.72587098605214 | 3.83260113686250 | 2.58217212245426 |
| D | 3.08346713279005 | 2.78005234904443 | 1.15108411706341 |
| O | 1.36567802043265 | 3.72828675104648 | 3.48638359269082 |
| D | 0.50879731695016 | 4.16050004295871 | 3.48765800258116 |
| D | 1.65631060691908 | 1.97085046075025 | 3.81587565051302 |
| O | 1.94877440323878 | 1.03805106580930 | 3.77440103837564 |
| D | 1.22541862337503 | 0.57321395380927 | 3.34211563634646 |
| D | 3.19224132532807 | 1.07977453194426 | 2.53875105019540 |
| O | 3.74029249672157 | 1.22882509451035 | 1.73491457367197 |
| D | 4.65040948834597 | 1.07637570958656 | 1.99942460018346 |

Geo #9

| | | | |
|---|---|---|---|
| C | 0.38728651384714 | -0.43204772625366 | 0.36164328614003 |
| O | -0.10567725568781 | -1.33599907114627 | -0.11012566584850 |
| D | 2.83132231123195 | -4.99181218409773 | -2.13862653580643 |
| O | 2.93448124822849 | -4.21590966198574 | -2.69422139995951 |
| D | 3.90609019277404 | -2.82643688605698 | -2.19368179390267 |
| D | 2.03631131529543 | -3.83666308237103 | -2.80346852168266 |
| O | 4.26422750836148 | -1.94272243934857 | -1.96273614076466 |
| D | 5.12446106796964 | -1.88651223430990 | -2.38485186641943 |
| D | 2.90406330373792 | -0.84728133700420 | -2.30483044329626 |
| O | 2.02224074031153 | -0.45422270067376 | -2.47790097999440 |
| D | 2.16079950277413 | 0.23347617054481 | -3.13332510898920 |
| D | 1.01506109484251 | -1.88330030449210 | -2.85436637063242 |
| O | 0.61507988349100 | -2.76957603554025 | -2.97418843876057 |
| D | -0.09574743717750 | -2.81099252726457 | -2.32932000008328 |